\documentclass[12pt]{iopart}

\usepackage{dcolumn}
\usepackage{bm}
\usepackage{ifpdf}
\usepackage{hyperref}
\usepackage{dcolumn}
\usepackage{bm}
\usepackage[spanish,english]{babel}
\usepackage{amsfonts}
\usepackage{amssymb}
\usepackage{graphicx}
\usepackage[latin1]{inputenc}

\usepackage{tikz}
\usetikzlibrary{decorations.pathmorphing}

\newcommand{\be}{\begin{equation}}
\newcommand{\en}{\end{equation}}
\newcommand{\bea}{\begin{eqnarray}}
\newcommand{\ena}{\end{eqnarray}}

\begin{document}

\title{Impact of curvature divergences on physical observers in a wormhole space-time with horizons}
\author{Gonzalo J. Olmo$^{1,2}$, D. Rubiera-Garcia$^3$, and A. Sanchez-Puente$^1$}
\address{$^1$Departamento de F\'{i}sica Te\'{o}rica and IFIC, Centro Mixto Universidad de
Valencia - CSIC. Universidad de Valencia, Burjassot-46100, Valencia, Spain}
\address{$^2$Departamento de F\'isica, Universidade Federal da
Para\'\i ba, 58051-900 Jo\~ao Pessoa, Para\'\i ba, Brazil}
\address{$^3$Instituto de Astrof\'isica e Ci\^encias do Espa\c{c}o, Universidade de Lisboa, Faculdade de Ci\^encias, Campo Grande, PT1749-016 Lisboa, Portugal}
\eads{\mailto{gonzalo.olmo@csic.es}, \mailto{drgarcia@fc.ul.pt}, \mailto{asanchez@ific.uv.es}}

\begin{abstract}
The impact of curvature divergences on physical observers in a black hole space-time which, nonetheless, is geodesically complete is investigated. This space-time is an exact solution of certain extensions of General Relativity coupled to Maxwell's electrodynamics and, roughly speaking, consists on two Reissner-Nordstr\"{o}m (or Schwarzschild or Minkowski) geometries connected by a spherical wormhole near the center.
We find that, despite the existence of infinite tidal forces, causal contact is never lost among the elements making up the observer. This suggests that curvature divergences may not be as pathological as traditionally thought.
\end{abstract}


\submitto{\CQG}

\maketitle

\section{Introduction}

Black holes are fascinating objects able to produce the largest known deformations of the causal structure of space-time. Their enormous gravitational pull generates trapped surfaces from which nothing can escape. In the framework of general relativity (GR), once a trapped surface is formed then a space-time singularity is unavoidable if certain reasonable conditions on the matter fields are satisfied \cite{Theorems1,Theorems2,Theorems3,Theorems4,Theorems5}. Formally, a singular space-time is characterized by the existence of past or future inextendible
null or time-like geodesics (geodesic incompleteness) \cite{Senovilla:2014gza,Geroch:1968ut,Wald:1984rg,Hawking:1973uf}. From a practical point of view, however, there is a widespread tendency to associate singularities with the presence of curvature divergences, which has led to numerous examples of nonsingular space-times\footnote{Note that, for singularity avoidance through bounded curvature scalars to occur, any of the assumptions of the singularity theorems has to be removed. Roughly speaking this has split the approaches into those violating the energy conditions (in the case of GR), or those considering theories extending GR, where violation of the energy conditions is not strictly needed.} using a variety of approaches \cite{RegularBHs}. This tendency is naturally justified by the intimate correlation existing between the blow up of curvature scalars and the inextendibility of geodesics as certain regions are approached. However, it has been known for a long time that both concepts are not equivalent for the characterization of space-time singularities \cite{Geroch:1968ut} (see also \cite{Curiel2009} for a discussion of this point). This means that, in principle, there could be space-times where the presence of pathologies in (some of) their curvature invariants  does not necessarily imply geodesic incompleteness, though explicit examples in physically consistent theories are hard to find.

In this sense, we have recently shown with an explicit example \cite{Olmo:2015dba,Olmo:2015bya} that black hole space-times with curvature divergences may exist which do not prevent the extension of null, time-like, or space-like geodesics to arbitrarily large values of their affine parameter. In other words, the correlation mentioned above between curvature divergences and geodesic incompleteness is explicitly broken. This occurs in a space-time which is essentially coincident with the Reissner-Nordstr\"{o}m (or Schwarzschild or Minkowski, depending on the choice of parameters) solution of GR everywhere except in a region close to the center, where a wormhole arises giving structure to the standard point-like singularity and connecting two identical copies of this classical geometry. Geodesic completeness accompanied by curvature divergences have also been recently found independently in models of quantum cosmology \cite{Singh:2011gp,Singh:2009mz} [see also \cite{JL}].

Our space-time turns out to be an exact solution of high-energy extensions of GR coupled to a spherically symmetric sourceless Maxwell field, whose properties have been studied in detail in a number of works \cite{or12,or13}. The gravitational Lagrangian of this theory might be motivated by well-established results of the theory of quantized fields in curved space-times \cite{QFT} and/or by Born-Infeld like extensions of GR, though formulated in the metric-affine (or Palatini) approach \cite{olmo}. This means that no a priori constraint on the relation between the metric and affine structures of the theory is imposed (for a pedagogic discussion of these concepts, see \cite{Zanelli,Olmo:2012yv}). The resulting scenario has some important features, such as the absence of higher-order field equations and ghost-like instabilities. On the other hand, since we deal with standard electric fields, the matter sector naturally satisfies the energy conditions. Accordingly, the theoretical framework can be regarded as physically consistent. This result puts forward that singularities can be resolved  in classical  geometric scenarios,  not requiring bounded curvature scalars or the invocation of quantum gravity effects for this purpose.

Though according to the formal classical criterion for space-time singularities one can say that the solutions found and discussed in \cite{or12} represent non-singular space-times, from a physical perspective it is important to determine if physical observers experience any pathological effect as they go through these regions with divergent curvature scalars. In fact, given that physical observers can be described in terms of congruences of geodesics, and that the evolution of these congruences depends on the components of the Riemann tensor, some of which are divergent, it is important to clarify if these observers can safely go through the wormhole, if they undergo some kind of deformation, or simply if they are destroyed in their transit. An in-depth exploration of this issue is thus necessary and motivates this work.

Before proceeding with the analysis, we would like to mention that the wormholes studied in this work are traversable in the sense that geodesics can go through them. Thus, we are not dealing with the traditional {\it traversable wormholes} of the literature [see \cite{wormholes} for a comprehensible overview of the topic], which are engineered solutions of the Einstein field equations designed to allow macroscopic objects to go through them and come back safely for whatever purposes (mainly for interstellar travel or as means to build a time machine). The model we are considering does allow solutions without event horizons (and thus traversable in the standard sense) and without curvature divergences at the throat, but represent microscopic entities not suitable for the transit of macroscopic spaceships. We are specially interested in the case in which massive geodesic observers can go through the wormhole and interact with the curvature divergences. These solutions present event horizons and, therefore, are not traditional {\it traversable wormholes}.  Nonetheless, understanding the impact of the transit on physical observers  is a question of theoretical (and maybe also practical) interest.

The content of this paper is organized as follows. In Sec. \ref{sec:geometry} we introduce the background geometry that shall be used through this paper and briefly describe its properties. In Sec. \ref{sec:Jacobis} the concepts of congruence of geodesics and volume elements are introduced, and subsequently applied to spherically symmetric wormholes in Sec. \ref{sec:spherical}. The corresponding effects for physical observers crossing the wormhole are discussed in Sec. \ref{sec:physical} and we finish in Sec. \ref{sec:conclusions} with some conclusions.

\section{Background geometry} \label{sec:geometry}

The background geometry we are going to study has been described in detail in \cite{Olmo:2015bya}. For completeness, we summarize here only those elements that are essential for the purposes of this work. For a more exhaustive presentation see \cite{Olmo:2015bya}. For convenience, we write the line element in the form

\begin{equation}\label{eq:ds2}
ds^2=-A(x)dt^2+\frac{1}{B(x)}dx^2+r^2(x)d\Omega^2 \ ,
\end{equation}
where the functions $A(x), B(x)$, and $r^2(x)$ are defined as

\begin{eqnarray}\label{eq:A}
A(x)&=& \frac{1}{\sigma_+}\left[1-\frac{r_S}{ r  }\frac{(1+\delta_1 G(r))}{\sigma_-^{1/2}}\right] \\
\delta_1&=& \frac{1}{2r_S}\sqrt{\frac{r_q^3}{l_\epsilon}} \\
\sigma_\pm&=&1\pm \frac{r_c^4}{r^4(x)} \\
B(x)&=& A(x)\sigma_+^2 \\
r^2(x)&=& \frac{x^2+\sqrt{x^4+4r_c^4}}{2} \label{eq:r(x)} \ ,
\end{eqnarray}
with the constant $r_c$ defined as $r_c=\sqrt{l_\epsilon r_q}$, with $l_\epsilon$ a length scale characterizing the high-curvature corrections in the gravity Lagrangian (which could be of order the Planck length $l_P=\sqrt{\frac{\hbar G}{c^3}}$), and $r_q^2=2G_N q^2$ a length scale associated to the electric charge, which together with the Schwarzschild radius $r_S \equiv 2M_0$ fully characterize the solution. The function $G(z)$, with $z=r/r_c$, is defined as

\begin{equation}
G(z)=-\frac{1}{\delta_c}+\frac{1}{2}\sqrt{z^4-1}\left[f_{3/4}(z)+f_{7/4}(z)\right] \ ,
\end{equation}
where $f_\lambda(z)={_2}F_1 [\frac{1}{2},\lambda,\frac{3}{2},1-z^4]$ is a hypergeometric function, and $\delta_c\approx 0.572069$ is a constant.

At scales $z\gg 1$, the space-time described by the line element (\ref{eq:ds2})  reduces to the standard Reissner-Nordstr\"om solution of GR, with $G(z)\approx -1/z$, $\sigma_\pm \approx 1$, $r^2(x)\approx x^2$, and

\begin{equation} \label{eq:RNsolution}
A(x)\approx 1-\frac{r_S}{ r  }+\frac{r_q^2}{2r^2} \ .
\end{equation}
Accordingly, the external horizon in this space-time is very close to the expectation from GR except for configurations with small values of the parameters $r_S$ and $r_q$ (microscopic black holes) \cite{or12}. However, the metric behavior close to the center, $x\to 0$, is rather different from the GR geometry. Defining the number of charges as $N_q=|q/e|$, where $e$ is the electron charge, and expanding the metric function is this region yields
\begin{equation}\label{eq:A_expansion}
\lim_{r\to r_c} A(x)\approx \frac{N_ql_P}{4N_c l_\epsilon}\frac{\left(\delta _1-\delta _c\right) }{\delta _1 \delta _c }\sqrt{\frac{r_c}{ r-r_c} }+\frac{1}{2}\left(1-\frac{N_ql_P}{N_c l_\epsilon}\right)+O\left(\sqrt{r-r_c}\right) \ ,
\end{equation}
which shows that the metric is finite at $r=r_c$ only for $\delta_1=\delta_c$, and diverges otherwise (recall that $l_P$ is the Planck length). We have introduced the constant $N_c\equiv \sqrt{2/\alpha_{em}}\approx 16.55$, where $\alpha_{em}$ is the electromagnetic fine structure constant. We note that the cases with $\delta_1=\delta_c$ are always free of curvature divergences (as can be directly checked, see \cite{or12}) and possess an event horizon if $N_q>N_c$ (here and from now on we set $l_P=l_\epsilon$). Those cases with $\delta_1 \neq \delta_c$ present curvature divergences over the sphere $r=r_c$, where ${R^\alpha}_{\beta\gamma\lambda}{R_\alpha}^{\beta\gamma\lambda} \sim(\delta_1-\delta_c)^2K_2/(r-r_c)^3+(\delta_1-\delta_c)K_1/(r-r_c)^{3/2}+ K_0$  (with the $K_i$ constants). Note that this divergence is much milder than the  $\sim 14r_q^4/r^8$ behavior of the Reissner-Nordstr\"om solution of GR. According to their macroscopic properties and number of horizons, we will refer to the cases with $\delta_1<\delta_c$ as Schwarzschild-like and to $\delta_1>\delta_c$ as Reissner-Nordstr\"{o}m-like. Note that the parameter $\delta_1$ represents a charge-to-mass ratio, which somehow justifies why the case $\delta_1<\delta_c$ is closer to a Schwarzschild black hole.

The fact that this solution has been derived assuming a sourceless electric field,  that the area function $r^2(x)$ reaches a minimum at $x=0$, and that the magnitude of the electric field at the surface $r=r_c$ is a universal quantity independent of $\delta_1$ (thus insensitive to the existence or not of curvature divergences) allows us to identify this geometry as a {\it geon} in Wheeler's original sense \cite{Wheeler}, namely, as a self-gravitating electric field with a wormhole structure \cite{W&M} (see \cite{wormholes} for a more detailed account on wormhole solutions). This implies that the coordinate $x\in [-\infty,+\infty]$, whereas $r(x)\ge r_c>0$. The presence of this finite-size wormhole structure modifies the space-time as compared to the standard Reissner-Nordstr\"om and Schwarzschild solutions of GR, in such a way that those time-like, null, and/or space-like geodesics that reach the wormhole can go through it and be extended to arbitrarily large values of their affine parameter, as shown in detail in \cite{Olmo:2015bya}. It thus constitutes an explicit example of geodesically complete space-time, no matter the behavior of curvature scalars.

\section{Classification and Jacobi fields \label{sec:Jacobis}}

The impact of curvature divergences on physical observers has been previously studied in the literature, leading to the establishment of certain criteria to estimate their strength and physical meaning. In this sense, the concept of {\it strong curvature singularity} \footnote{Given the characteristics and scope of our work, from now on we will replace the traditional term {\it curvature singularity} by {\it curvature divergence}  to emphasize that curvature divergences need not imply space-time singularities.} was originally introduced by Ellis and Schmidt \cite{Ellis}, who identified a strong curvature divergence by the property that {\it all} objects falling into it are crushed to zero volume, no matter what their physical features are. This statement captures the notion that space-time singularities are strictly geometric phenomena, not related to specific properties of the matter. This intuitive definition was given precise mathematical form by Tipler \cite{Tipler} and was further developed by Clarke and Krolak \cite{CK}. Some refinements of the initial characterization were introduced later on to include in the strong group some (originally weak) cases in which the volume remains finite but the body undergoes unacceptably large deformations (see \cite{Nolan, Ori} for details).

The key idea behind the above classification is to somehow idealize a body as a set of points following geodesics of the background metric. One then studies the evolution of the separation between nearby geodesics as the singularity is approached to determine its impact on the body \cite{Tipler77}.
In mathematical terms, one considers a congruence of geodesics labeled by means of two parameters $x^\mu=x^\mu(\lambda,\xi)$, where $\lambda$ represents the affine parameter along a given geodesic, and $\xi$ serves to identify the different geodesics on the congruence. For a given geodesic, the tangent vector is $u^\mu\equiv \partial x^\mu/\partial \lambda $ (with constant $\xi$). The separation between nearby geodesics (at given $\lambda$) is measured by the Jacobi vector fields $Z^\mu\equiv \partial x^\mu/\partial \xi$, which satisfy the geodesic deviation equation

\begin{equation}\label{eq:Jacobi}
\frac{D^2 Z^\alpha}{d\lambda^2}+{R^\alpha}_{\beta\mu\nu}u^\beta Z^\mu u^\nu=0 \ ,
\end{equation}
where $D Z^\alpha/d\lambda\equiv u^\kappa \nabla_\kappa Z^\alpha=Z^\beta \nabla_\beta u^\alpha$ (see, for instance, chapter 11 of \cite{MTW}). Using an adapted orthonormal tetrad  parallel transported along the congruence and with the (normalized) tangent vector $u^\mu \partial_\mu$ defining the basis vector $e_0$, we consider only those separation vectors orthogonal to $e_0$, i.e., those contained in the subspace spanned by the basis vectors $\{e_1,e_2,e_3\}$ (see chapter 4 of \cite{Hawking:1973uf} for details). These vectors can be written as $Z=Z^a e_a$, with components $Z^a$ ($a=1,2,3$). Given the second-order character of Eq.(\ref{eq:Jacobi}), it follows that there are six independent Jacobi fields along a given geodesic depending on the values of $Z^a$ and $D Z^a/d\lambda$ at some point $\lambda_i$. If the $Z^a(\lambda_i)$ are not all zero, then the linearity of Eq.(\ref{eq:Jacobi}) allows to express  the components $Z^a(\lambda)$ at any $\lambda$ in terms of their values at  $\lambda_i$  by

\begin{equation}
Z^a(\lambda)={A^a}_b(\lambda) Z^b(\lambda_i),
\end{equation}
with ${A^a}_b(\lambda)$ a $3\times 3$ matrix which is the identity matrix at $\lambda=\lambda_i$. If the $Z^a(\lambda_i)$ are all zero at $\lambda_i$, then one can write

\begin{equation}
Z^a(\lambda)={\mathcal{A}^a}_b(\lambda) \left.\frac{D Z^b}{d\lambda}\right|_{\lambda=\lambda_i},
\end{equation}
with ${\mathcal{A}^a}_b(\lambda)$  a $3\times 3$ matrix which vanishes at $\lambda_i$. In the latter case, the Jacobi fields represent the separation of neighbouring geodesics that meet at $\lambda_i$.

With three linearly independent solutions of (\ref{eq:Jacobi}), $Z_{(i)}=Z^a_{(i)}e_a$ ($i=1,2,3$), one can define a volume element (via a three-form) given by

\begin{equation}
V(\lambda)= \det[Z_{(1)}^a, Z_{(2)}^b, Z_{(3)}^c].
\end{equation}
This volume element, which is independent of the orthonormal basis chosen with $e_0=u$, can be related to the determinant of the matrix ${A^a}_b(\lambda)$ (or ${\mathcal{A}^a}_b(\lambda)$ if $Z^a(\lambda_i)=0$) and the details of the initial configuration at $\lambda_i$ as

\begin{equation}
V(\lambda)=\det|A(\lambda)|V(\lambda_i)
\end{equation}
(or with $\det|\mathcal{A}(\lambda)|$ if $Z^a(\lambda_i)=0$). This puts forward that the details of the initial configuration (either the volume defined by the $Z^a_{(j)}(\lambda_i)$ or by the $\left.\frac{DZ^a_{(j)}}{d\lambda}\right|_{\lambda_i}$) are not essential to determine the strength of a singularity. Only the functional dependence of ${A^a}_b(\lambda)$ (or ${\mathcal{A}^a}_b(\lambda)$ if $Z^a(\lambda_i)=0$) is necessary. Recall that, according to Clarke and Krolak \cite{CK} (see also \cite{Tipler77}), a strong singularity occurs when

\begin{equation}
\lim_{\lambda \rightarrow 0} V(\lambda)=0
\end{equation}
with $\lambda=0$ representing the arrival to the singularity.

\section{Spaces with spherical symmetry} \label{sec:spherical}

In spherically symmetric space-times, Nolan \cite{Nolan} provided a transparent analysis of the strength of singularities following the more general (and abstract) approach of Clarke and Krolak \cite{CK}. Following \cite{Nolan}, the Jacobi fields can be taken as

\begin{eqnarray}\label{eq:Jacobi_1}
Z_{(1)}&=& B(\lambda)({u^x}/{A},A u^t,0,0) \\
Z_{(2)}&=& (0,0,P(\lambda),0) \\
Z_{(3)}&=& (0,0,0,Q(\lambda)/\sin\theta) \,
\end{eqnarray}
which are orthogonal to the time-like, radial geodesic tangent vector $u^\mu=(u^t,u^x,0,0)$, where $u^t\equiv dt/d\lambda=E/A$, with $E=$ constant representing the total energy per unit mass for time-like geodesics, and $u^x\equiv dx/d\lambda$ is such that

\begin{equation}
\left(\frac{dx}{d\lambda}\right)^2=\sigma_+^2(E^2-\kappa A),
\end{equation}
with $\kappa=1$ for time-like geodesics and $\kappa=0$ in the null case (see \cite{Olmo:2015bya} for more details, including the case with nonzero angular momentum).

The functions $B(\lambda)$, $P(\lambda),$ and $Q(\lambda)$ must be determined via the geodesic deviation equation (\ref{eq:Jacobi}). One finds that $P(\lambda)$ and $Q(\lambda)$ admit identical solutions of the form

\begin{equation}\label{eq:P}
P(\lambda)=P_0+C\int \frac{d\lambda}{r^2(\lambda)} \ ,
\end{equation}
whereas for $B(\lambda)$ one finds the following equation

\begin{equation}\label{eq:a}
B_{\lambda\lambda}+\frac{A_{yy}}{2}B(\lambda)=0 \ ,
\end{equation}
where $y(x)=\int dx/\sigma_+$ and $x=x(\lambda)$ is determined by integrating $dx/d\lambda$, which can be approximated near the wormhole  as $x(\lambda)\approx \pm (9a \lambda^2)^{1/3}$ (for outgoing/ingoing geodesics), where $a=\kappa\frac{ N_q (\delta_c-\delta_1)}{2N_c\delta_c \delta_1 }$. (Note that we are considering the Schwarzschild-like case $\delta_1<\delta_c$ because this is the only wormhole configuration in which time-like geodesics can reach the divergence \cite{Olmo:2015bya}). The function $r^2(\lambda)$ can also be written as $r^2(\lambda)\approx r_c^2+x^2/2$, $\sigma_+\approx 2$ when $x\to 0$, and $A(x)\approx -a/|x|$. As a result, we find that near the singularity we can approximate (\ref{eq:a}) as

\begin{equation}\label{eq:a2}
B_{\lambda\lambda}-\frac{4}{9\lambda^2}B(\lambda)=0 \ ,
\end{equation}
which admits an exact solution. Imposing the standard initial condition that all Jacobi fields vanish at the initial point $\lambda_i$, we find the following solutions in the neighborhood of the singularity (here $C_1, C_2,$ and $C_3$ are arbitrary constants)

\begin{eqnarray}
B(\lambda)&\approx& C_1\left(\frac{1}{|\lambda|^{1/3}}-\frac{|\lambda|^{4/3}}{|\lambda_i|^{5/3}} \right)\\
P(\lambda)&\approx& C_2 (\lambda-\lambda_i) \\
Q(\lambda)&\approx& C_3 (\lambda-\lambda_i) \ .
\end{eqnarray}
The resulting volume is given by the product \cite{Nolan}

\begin{equation}
V(\lambda)=|B(\lambda) P(\lambda) Q(\lambda)| r^2(\lambda),
\end{equation}
which behaves as $V(\lambda)\sim 1/\lambda^{1/3} $ as the singularity is approached at $\lambda=0$. Rather than vanishing, this volume diverges due to the behavior of the {\it radial} Jacobi field $Z_{(1)}$, whose modulus grows without bound as $\lambda\to 0$. One can easily verify that the behavior of $Z_{(1)}$ here is identical to that found in a standard Schwarzschild black hole. In that case, we have that $A(r)=1-r_S/r$  and $dr/d\lambda= \pm\sqrt{r_S/r+E^2-1}$, which near $r\to 0$ turns (\ref{eq:a})  into exactly the same form as (\ref{eq:a2}). The angular part, however, is clearly different because near the Schwarzschild singularity we have $r(\lambda)\approx \left(\frac{9r_S}{4}\right)^{1/3}\lambda^{2/3}$, which leads to $P(\lambda)\approx \tilde{C}(|\lambda_i|^{-1/3}-|\lambda|^{-1/3})$, with $\tilde{C}$ another integration constant. As a result, the product $P(\lambda) Q(\lambda) r^2(\lambda)\propto \lambda^{2/3}$ and $V(\lambda)\sim \lambda^{1/3}$, which vanishes as $\lambda\to 0$, thus signalling the presence of a strong divergence according to Tipler's criterium \cite{Tipler}. The angular part, therefore, makes all the difference between the usual Schwarzschild curvature divergence and the divergence of our wormhole in the Schwarzschild-like configuration as far as time-like geodesics are concerned\footnote{We note here that the area element carried by null geodesics in the Schwarzschild space-time is well behaved in GR as well as in our context. See \cite{Nolan} for a discussion of the Jacobi fields in the case of null geodesics. }. In the Schwarzschild case, all geodesics that meet at $\lambda_i$ converge again at the center. In our case, the finite radius of the wormhole prevents this convergence, and a finite angular separation between geodesics remains constant as the divergence is approached.  According to Ori \cite{Ori} and Nolan \cite{Nolan:2000rn} the case $\lim_{\lambda\to 0} V\to \infty$ could be regarded as {\it deformationally} strong.

\section{Physical interpretation and implications} \label{sec:physical}

We have just seen that the curvature divergence in the Schwarzschild black hole and in the Schwarzschild-like configuration of our wormhole can be reinterpreted in terms of the collapse ($V\to 0$) or divergence ($V\to \infty$) of a volume element transported by physical observers. In the standard Schwarzschild case, the fact that ingoing geodesics terminate at $r=0$ and that all the elements in a congruence of radial time-like geodesics converge at this point is a signal of the destructive and pathological nature of this region. In the wormhole case, however, all geodesics are complete and the fact that the volume defined by a congruence of time-like geodesics diverges at the throat deserves further scrutiny to understand its physical implications. In fact, the analysis in terms of Jacobi fields seems to have simply replaced a divergence in curvature scalars by a divergence in a certain volume element. In this section we will try to shed some light on this issue by exploring the definition of this volume element from the perspective of a freely falling observer.

To proceed, we find it useful to write the line element of our space-time in coordinates adapted to a freely falling family of observers with a reference energy $E$ (and zero angular momentum for simplicity). We can thus define a new time coordinate that coincides with the tangent vector of time-like observers according to $\partial_\lambda=u^t\partial_t+u^y\partial_y$, where $u^y\equiv dy/d\lambda=\pm \sqrt{E^2-A}$ has been written in terms of the rescaled coordinate $y=\int^x dx/\sigma_+$ for simplicity. We could also propose a {\it radial} coordinate $\partial_{\tilde{\xi}}$ orthogonal to $\partial_\lambda$ and to the spherical sector in the form $\partial_{\tilde{\xi}}=(u^y/A)\partial_t+A u^t\partial_y$. This vector has unit norm and points in the same {\it radial} direction as the Jacobi field $Z_{(1)}$ given in (\ref{eq:Jacobi_1}). Unfortunately, this choice leads to $[\partial_\lambda,\partial_{\tilde{\xi}}]\neq 0$ and, therefore, does not define a coordinate basis. One can verify, however, that $\partial_\xi=u^y[(u^y/A)\partial_t+A u^t\partial_y]$ does define a good coordinate basis with $[\partial_\lambda,\partial_\xi]= 0$. It is possible to find an explicit form for the change of coordinates:

\begin{equation}
 \lambda(y,t) = -E t+\int_0^y \frac{u^y}{A} dy^\prime  \ , \  \xi(y,t) = -t+\int_0^y \frac{u^t}{u^y} dy^\prime
\end{equation}
It is also possible to get the old coordinate $y$ in terms of $\lambda$ and $\xi$ inverting the following relation:

\begin{equation}
\lambda-E \xi = \int_0^y \frac{1}{u^y(y^\prime)} dy^\prime
\end{equation}
With this choice of coordinates, the wormhole throat, $x=0$, is found at the hypersurface $\lambda-E \xi=0$. The line element (\ref{eq:ds2}) (with $dy^2=dx^2/\sigma_+^2$) turns into

\begin{equation}\label{eq:FF}
ds^2=-d\lambda^2+(u^y)^2d\xi^2+r^2(\lambda,\xi)d\Omega^2 \ .
\end{equation}
It is worth noting that this line element is intimately related to the Jacobi fields introduced before. In fact, a geodesic deviation vector $Z^\mu$ connecting two nearby geodesics $x_1^\mu(\lambda)$ and $x_2^\mu(\lambda)$ occupying the locations $\xi=\xi_1$ and $\xi=\xi_2$ in the same congruence is defined as $x_2^\mu(\lambda)-x_1^\mu(\lambda)=Z^\mu(\lambda)\Delta \xi$ in the limit $\Delta\xi\to 0$, i.e., $Z^\mu=\partial x^\mu(\lambda,\xi)/\partial\xi$. For infinitesimally close geodesics, $dx^\mu=  Z^\mu_{(i)}(\lambda) d\xi^{(i)}$, with the index $(i)$ denoting the independent spatial directions. The spatial distance between nearby geodesics at a given $\lambda$ is thus given by $ds^2=g_{\mu\nu}Z^\mu_{(i)}Z^\nu_{(j)}d\xi^{(i)}d\xi^{(j)}$, with $g_{\mu\nu}Z^\mu_{(i)}Z^\nu_{(j)}$ representing the square of the norm of the Jacobi fields. The volume element defined by the Jacobi fields is thus intimately related to the {\it infinitesimal} volume element defined by the spatial coordinates of a freely falling observer.

Near the wormhole in the Schwarzschild-like case, $(u^y)^2$ can be approximated as $(u^y)^2 \simeq a/|x| \simeq (\frac{3}{a} |\lambda-E\xi|)^{-\frac{2}{3}}$, which turns (\ref{eq:FF}) into

\begin{equation}\label{eq:FF_WH}
ds^2\approx-d\lambda^2+\left(\frac{3}{a} |\lambda-E\xi|\right)^{-2/3}d\xi^2\ .
\end{equation}
This expression puts forward that, as the wormhole throat is approached,
the physical spatial distance between any two infinitesimally nearby radial geodesics diverges: $dl_{Phys}=\left(\frac{3}{a} |\lambda-E\xi|\right)^{-1/3}d\xi$. However, for any finite {\it comoving} separation $l_\xi\equiv \xi_1-\xi_0$, the physical spatial distance $l_{Phys}\equiv \int |u^y| d\xi$ is given by
\begin{equation}\label{eq:lphys}
l_{Phys}\approx \left(\frac{a}{3}\right)^{1/3}\frac{1}{E}\left||\lambda-E \xi_0|^{2/3}-|\lambda-E \xi_1|^{2/3}\right| \ ,
\end{equation}
which always yields a finite physical length. This result is very important and puts forward that the infinite stretching of the {\it infinitesimal} spatial distance in the radial direction is at the root of the divergent behavior of the volume element carried by the Jacobi fields discussed above in Sec.\ref{sec:spherical} (where $\xi=0$ was chosen as the fiducial geodesic). Now, given that any finite comoving separation in the radial direction remains finite and that the angular sector is well-behaved at the wormhole throat\footnote{ Recall that in the Schwarzschild solution of GR the singularity is regarded as strong because the angular sector rapidly collapses to zero, making the infinitesimal volume carried by the congruence to vanish despite the divergence in the radial sector. In the Schwarzschild-like solution studied here, the angular sector tends to a constant as the curvature divergence is approached, which can be interpreted as the fact that the congruence defines a nonzero finite area at the throat.}, any finite (non infinitesimal) volume crossing the wormhole should remain finite at all times. This suggests that the different elements of an extended body that goes through the wormhole should remain in causal contact during the transit. A detailed calculation is thus necessary.  

From the above analysis one finds that infinitesimally nearby geodesics are infinitely stretched in the radial direction in a process that, however, is reversed as soon as  the wormhole is crossed. A natural question to ask, therefore, is if this process of {\it spaghettization} (experienced as the wormhole is approached) followed by an identical contraction (as the wormhole is left behind) has any physical impact on objects crossing the wormhole. In particular, if the constituents making up an object that reaches the wormhole lose causal contact because of the {\it spaghettization} process, then the interactions that keep the object cohesioned would no longer be effective, which would result in disruption or disintegration of the body.  In that case, one should necessarily conclude that the object has been destroyed due to the presence of a strong curvature divergence.

\begin{figure}[h]
\begin{center}
\includegraphics[width=0.5\textwidth]{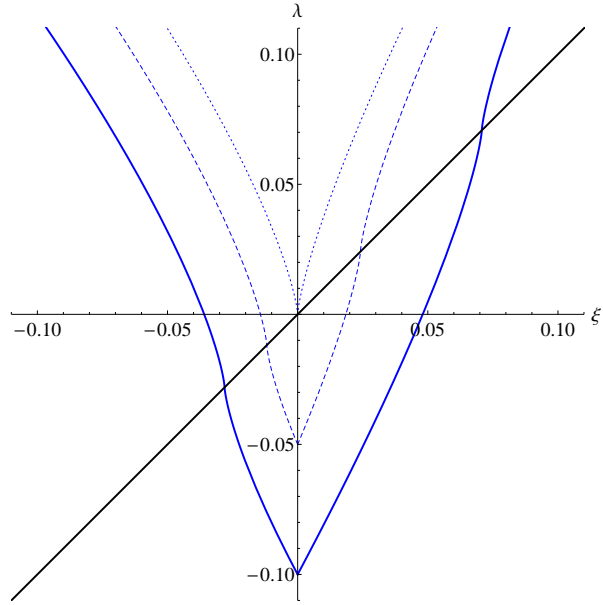}
\caption{Trajectories of light rays emitted by a freely falling observer from $\xi=0$ at different times shortly before reaching the wormhole throat in the Schwarzschild-like configuration. The rays going to the left/right represent ingoing/outgoing null geodesics. Given that the observer is inside an event horizon, both ingoing and outgoing light rays end up hitting the wormhole. The wormhole throat is located at the oblique (solid black) line $\lambda - E\xi= 0$ (in the plot $E=1, a=3$). }\label{fig:light-cones}
\end{center}
\end{figure}

To explore this aspect, consider the propagation of radial null rays according to (\ref{eq:FF_WH}). Since for null rays $ds^2=0$, we have that photon paths satisfy

\begin{equation}
\frac{d\xi}{d\lambda}= \pm \left|\frac{3}{a} (\lambda-E\xi)\right|^{1/3}.
\end{equation}
Numerical integration of these equations leads to the graphic representation of  light cones in Fig.\ref{fig:light-cones}.

\begin{figure}[h]
\begin{center}
\includegraphics[width=0.5\textwidth]{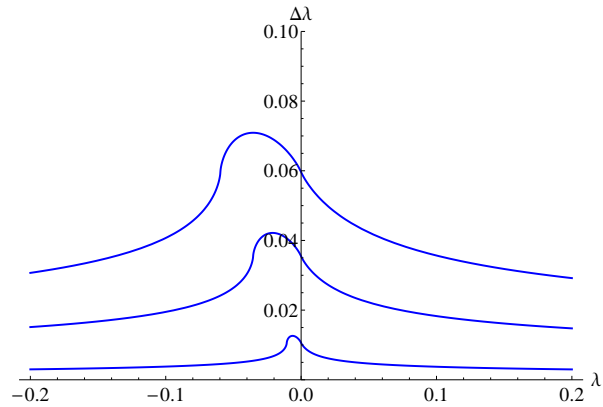}\label{TimeLightRays}
\caption{Representation of the proper time $\Delta \lambda$ that a light ray takes in a round trip from a fiducial geodesic at $\xi=0$ to another separated radially by comoving distance $\xi =0.01 r_c,0.005r_c,0.001r_c$, versus the value of the proper time $\lambda$ at which the light ray was sent. At $\lambda=0$ the geodesic encounters the curvature divergence. Light rays sent soon before reaching the wormhole will encounter the divergence on their way, causing an additional delay (the ``bumps" in the plot) in travelling time. This confirms that the travelling time is finite at all moments and tends to 0 as the comoving distance tends to 0. }\label{fig:TimeLightRays}
\end{center}
\end{figure}

As can be seen, physical observers near the wormhole are in causal contact with their neighborhood despite the infinite spatial stretching and contraction in the radial direction experienced as $\lambda\to E\xi$. Any nearby geodesic with $\xi\neq 0$ can be reached by a light ray in a finite (proper) time (see Fig.\ref{fig:TimeLightRays}) and, therefore, the interactions among the constituents of a body going through the wormhole are preserved. We must thus conclude that physical observers do not experience any dramatic effect as they go through $\lambda=0$, where the curvature divergence is located and the infinitesimal spatial volume diverges. Therefore, the existence of a curvature divergence seems to have very little physical impact if any on objects with a finite volume. According to this result, the application of the standard classification criteria for the strength of curvature divergences in the case of having a divergent volume element should be handled with care, paying special attention to the preservation of causal contact as a new source of useful information.

\section{Summary and conclusions} \label{sec:conclusions}

In this work we have studied some aspects of geodesic congruences to  explore the effects of curvature divergences on physical observers in a geodesically complete space-time with unbounded curvature scalars. Following previous analyses in the literature, we have investigated the behavior of the (infinitesimal) volume element carried by freely falling observers using a Jacobi field approach. This analysis has focused on the Schwarzschild-like case of the background geometry, which is the only one in which time-like observers can effectively go through the troublesome region (recall that, even in GR, time-like observers cannot reach the central divergence of the Reissner-Nordstr\"{o}m solution). We have found that this volume element diverges due to an infinite stretching experienced by infinitesimally separated radial geodesics at the wormhole throat, where curvature scalars blow up. This stretching is followed by an identical contraction once the throat is crossed.  The divergence of the volume element contrasts with the standard Schwarzschild picture, in which it goes to zero due to the rapid collapse of the angular directions, even though there is an infinite radial stretching identical to that found here in the wormhole case.

We have then shown that despite the infinite stretching in the radial direction of the spatial distance between infinitesimally nearby time-like geodesics, causal contact among them is never lost, which guarantees the effective transmission of interactions among the constituents of the body (see Figs.\ref{fig:light-cones} and \ref{fig:TimeLightRays}). Moreover, the physical spatial distance between non-infinitesimally separated time-like geodesics is always finite, as shown in (\ref{eq:lphys}). We thus conclude that physical observers do not perceive any dramatic sign of destruction as extended  objects cross the wormhole. The existence of curvature divergences in the space-times considered here, therefore, does not seem to cause any pathological effects on physical observers (either represented by individual geodesics or by congruences). This result indicates that the criteria used in the literature to classify the strength of curvature divergences when the Jacobi volume element diverges should be applied taking into account also the role of causality.

The analysis presented here as well as in \cite{Olmo:2015bya} has focused entirely on classical geometrical aspects of gravitation and physical observers represented by geodesics or congruences of geodesics. However, in order to better understand the impact of curvature divergences on systems with quantum properties further research is mandatory. A preliminary analysis of wave propagation in this context has been recently presented in \cite{Olmo:2015dba}, finding that the propagation is smooth despite the existence of divergent effective potentials related with the geometric divergences.

\section*{Acknowledgments}

G.J.O. is supported by a Ramon y Cajal contract, the Spanish grants FIS2014-57387-C3-1-P and FIS2011-29813-C02-02 from MINECO, the grants i-LINK0780 and i-COOPB20105 of the Spanish Research Council (CSIC). D. R.-G. is funded by the Funda\c{c}\~ao para a Ci\^encia e a Tecnologia (FCT) postdoctoral fellowship No.~SFRH/BPD/102958/2014, the FCT research grant UID/FIS/04434/2013, and the NSFC (Chinese agency) grant No.~11450110403. The authors also acknowledge support from CNPq (Brazilian agency) grant No. 301137/2014-5.

\end{document}